\documentstyle[prc,preprint,aps]{revtex}
\def\bea {\begin{eqnarray}}
\def\eea {\end{eqnarray}}
\def\be {\begin{equation}}
\def\ee {\end{equation}}
\def\ben{\begin{enumerate}}
\def\een{\end{enumerate}}
\def\bi{\begin{itemize}}
\def\ei{\end{itemize}}

\def\etal{{\it et al.}}

\def\pl {Phys. Lett.\ }
\def\pr {Phys. Rev.\ }
\def\np {Nucl. Phys.\ }

\begin{document} 
\tighten
\draft 
\preprint{ } 
\title{Shell-model calculations of neutrino scattering from
$^{12}$C }
\author{ 
A.C. Hayes and I.S. Towner\footnote{Present address: Department of Physics,
Queen's University, Kingston, Ontario K7L 3N6, Canada}}
\address{ 
Theoretical Division, Los Alamos National Laboratory,
Los Alamos, NM 87545} 
\date{\today} 
\maketitle
\begin{abstract} 
Neutrino reaction cross-sections,
$(\nu_\mu,\mu^-)$, $(\nu_e,e^-)$, $\mu$-capture  
and photoabsorption rates
on $^{12}$C are computed within a large-basis shell-model framework,
which included excitations up to $4\hbar\omega$. 
When ground-state correlations are included with an 
open $p$-shell the predictions of the 
calculations are in reasonable agreement with most of
the experimental results for these reactions.
Woods-Saxon radial wave functions are used, with their asymptotic
forms matched to the experimental separation energies for bound
states, and matched to a binding energy of 0.01 MeV for unbound
states.  For comparison purposes, some results are given for harmonic
oscillator radial functions.  Closest
agreement between theory and experiment 
is achieved with unrestricted shell-model configurations and        
Woods-Saxon radial functions.
We obtain for the neutrino-absorption inclusive cross sections:
$\overline{\sigma} = 13.8 \times 10^{-40}$ cm$^2$
for the $(\nu_{\mu},\mu^{-})$ decay-in-flight flux
in agreement with the LSND datum of
$(12.4 \pm 1.8) \times 10^{-40}$ cm$^2$; and
$\overline{\sigma} = 12.5 \times 10^{-42}$ cm$^2$
for the $(\nu_{e},e^{-})$ decay-at-rest flux,
less than the experimental result of
$(14.4 \pm 1.2) \times 10^{-42}$ cm$^2$. 
\end{abstract} 


\narrowtext 

\section{Introduction}
One of the key neutrino-nucleus reactions measured at the 
two accelerator based  neutrino-oscillations searches,
LSND at the LANSCE 
facility in Los Alamos and KARMEN at the ISIS facility  
in the U.K., is the scattering of neutrinos  
from carbon in the liquid scintillation detector.
The neutrino source at both experiments comes from the 
decay of pions produced 
in the beam stop. The vast majority of pions decay at 
rest, and the electron neutrinos thus produced 
have enough energy to cause a nuclear charge-exchange 
reaction on carbon,
$^{12}$C($\nu_e,e^-)^{12}$N. At LSND 3.4$\%$ of the 
pions decay in flight,
producing muon neutrinos of sufficient energy to 
interact via the reaction $^{12}$C($\nu_\mu,\mu^-)^{12}$N.
The signal for
the oscillation of decay-in-flight (DIF) muon 
neutrinos ($\nu_\mu\rightarrow \nu_e$) at LSND is the appearance
of high energy electrons from the $\nu_eC\rightarrow Ne^-$ reaction.
Extracting oscillation parameters from this search requires
knowledge of the expected cross-section. In addition, the measured
$\nu_\mu$C$\rightarrow$N$\mu^-$ cross-section
acts as a test of the $\nu_\mu$ DIF flux and of the detector efficiency.
The KARMEN experiment also has the efficiency to measure the
$\bar{\nu}C\rightarrow B\mu^+$ reaction.
At both LSND and KARMEN the inclusive $\nu_e$C$\rightarrow$X$e^-$ and
the exclusive $\nu_e$C$\rightarrow^{12}$N$_{g.s.}e^-$ cross-sections
are measured for the electron neutrinos from the decay of the 
pion at rest.
The flux of the decay-at-rest (DAR) electron neutrinos
is
given by the Michel spectrum, and
the $\nu_e$C cross sections provide a strong 
constraint on the nuclear-structure
models
used for carbon.

The inclusive $^{12}$C($\nu_\mu,\mu^-$)X cross section 
for the DIF $\nu_\mu$ flux at LSND has been measured.
The first calculations for this cross section were carried out by the
the Caltech group\cite{Ko95} using a 
continuum Random Phase Approximation (RPA) model.
The calculated cross sections over-estimated 
experiment by almost a factor of
two,
and suggested that the measured cross sections may be inconsistent
with other observables for $^{12}$C.
The most recent version of this work\cite{Ko99}
allows for a partial occupancy of the nuclear subshells
in the continuum RPA, a feature that brings
theory and experiment closer together, but still does not remove the
discrepancy fully.
The other key observables that need to be considered
are  $\mu$-capture, the DAR
$^{12}$C($\nu_e,e^-)X$,
$ (e,e')$, photo-absorption, and
$\beta$-decay. These different probes involve different energy and
momentum transfers and, thus, constrain different 
aspects of the calculations.
The disagreement between theory and experiment 
for the DIF $\nu_\mu$C cross section prompted
new
calculations\cite{Auerbach,oset}
in an effort to uncover possible shortcomings of the RPA calculations.
In the present work we perform a series of shell-model 
based calculations, which
 include excitations up to $4\hbar\omega$ for 
$^{12}$C, $^{12}$N and $^{12}$B, each involving 
different assumptions and approximations.

The DAR neutrino spectrum involves neutrino 
energies $0-52$ MeV, with an average
neutrino energy
$E_\nu\sim$ 32 MeV. The $Q$-value for the charge-exchange 
reaction ($\nu_e,e^-$)
on $^{12}$C is about 17 MeV.
The DAR inclusive cross-section
is then dominated by low multipoles (1$^+$, 1$^-$, 2$^-$)
and by excitation of the giant dipole resonances. 
In the case of the  $\nu_\mu$C cross section, the DIF muon neutrino
flux involves an  average neutrino 
energy of about $E_\nu\sim 150 $ MeV, but
the flux is finite up to  $E_\nu\sim 250$ MeV.
The $Q$-value for the ($\nu_\mu,\mu^-$)
reaction
is close to 123 MeV.
Calculations for this cross section need to 
include both a good description
of the giant resonance 
region  ($E_x\sim 15-40 $ MeV of excitation in $^{12}$C)
and of higher excitation energy regions
 (up to $80-100$ MeV).
Furthermore,
all multipoles $\lambda\sim 0-5$ make significant contributions to the
inclusive cross section.

\section{General Nuclear Structure Considerations}
\label{s:GNSC}

In the simplest shell model,  $^{12}$C consists of four 
neutrons and four protons
in the $p$-shell outside a closed $^4$He core. Excited states reached
by the $(\nu,\ell^-$) reactions are simple particle-hole
states built on the ground state. However, the structure of
both the $^{12}$C$_{g.s.}$ and the 
continuum states of $^{12}$N also involve
more sophisticated configurations and configuration mixing.
There is a limit on the size of the model space that can be included
in any calculation, so that calculated cross sections 
necessarily involve some level of approximation.
In the case of the neutrino
reactions on carbon  the approximations that are
most likely to affect the predicted cross sections are
$(a)$ the $^{12}$C ground state $p$-shell structure,
$(b)$ the treatment of ground state correlations beyond the $p$-shell,
$(c)$ the model space truncation, especially for the final states,
$(d)$ configuration mixing in the final states, and
$(e)$ the nuclear radial wave functions.
In this paper we examine the effect of each of these on the predicted
cross section.

An approximation that is inherent in the continuum RPA 
calculations of Kolbe \etal \cite{Ko95}
is the restriction of the $^{12}$C ground state to 
a closed $0p_{3/2}$ shell.
This approximation is necessary to
build a spectrum of particle-hole states representing the excitations
in $^{12}$N and to evaluate their transition cross sections in the
RPA.  That this approximation is
very poor for the lowest positive-parity states of configuration
$(p_{3/2}^{-1},p_{1/2})$ is well known and recognized in the 
calculations of Kolbe \etal \cite{Ko95}, who reduce their 
calculated cross section to
the $^{12}$N$_{g.s.}$ by a factor of four.  
A smaller reduction factor of order 1.5 is obtained when this
work is extended \cite{Ko99} to include partial occupancies
for the $p_{1/2}$ subshell.
An additional reduction of the cross section
to the other low-lying positive parity states, 
particularly to the 2$^+$ state at 0.96 MeV of excitation in $^{12}$N,
should also be included.
It is unclear whether any further correction is required for
particle-hole states of energy $1 \hbar \omega$, $2 \hbar \omega$ 
$\ldots$ above these lowest positive-parity states in $^{12}$N. 
Kolbe \etal \  have argued against additional 
suppression factors, other than the one
that has been applied to the GT transition to the $^{12}$N
ground state.
 
In the case of particle-hole excitations out of 
the $p$-shell, the $p$-shell structure of the
ground state does not affect the total sum-rule for 
a given operator, which is  determined
by the number of particles in the $p$-shell.
The total spin independent $\Delta S = 0$ multipole 
strength is independent of the $p$-shell structure of the ground state.
However, the total spin-dependent $\Delta S = 1$ strength 
is distributed differently over
the different spin-multipoles for different $p$-shell 
ground states \cite{OL93,ZA99}. Thus, for  flux-averaged neutrino 
reactions the predicted cross section  
will, in general,  be model dependent,
reflecting the model ground state spin structure.

A reasonable approximation for the structure of $^{12}$C is  the
$p$-shell equivalent of three $\alpha$-particles.  This corresponds to  
an $L=0$ $S=0$ ground state,  with good SU(4) symmetry, 
$[\widetilde{444}]$.
The Cohen-Kurath interaction\cite{CK65} predicts 
that this state makes up $78\%$ of the
$^{12}$C$_{g.s.}$ wave function. 
In contrast, the closed $p_{3/2}$-state contains $16\%$ $S=0$ and
$6\% [\widetilde{444}]$ symmetry, Table \ref{t:tab0}. 
Thus, the spin response for the closed $p_{3/2}$ state 
will differ from that of  the Cohen-Kurath ground state.
Determining the degree to which this model dependence is 
reflected in the predicted neutrino cross sections
is a main aim of the
present work.

Proton threshold in $^{12}$N is at 0.601 MeV, so that all 
levels but the ground state are unstable. States up to about
8 MeV of excitation in $^{12}$N have been 
studied via the $^{12}$C$(p,n)$ and $(n,p)$ charge exchange reactions. 
The structure of these states can be
understood largely within a (0+1)$\hbar\omega$ shell-model 
calculation, although the predictions of these Tamm-Dancoff
calculations
overestimate the charge exchange cross sections by about  
a factor of two. Little detailed information exists for
higher energy regions of the $^{12}$N continuum.

The unbound nature of the states in $^{12}$N is in 
strong contrast with the deeply bound $p$-shell particles in
the $^{12}$C ground state. Thus,  there will be a 
strong lack of overlap between the initial- and final-state
radial
functions. The continuum RPA calculations of Kolbe \etal \cite{Ko95}  
explicitly treat the continuum nature of the excited states
of $^{12}$N. In contrast, the present shell model 
calculations are discrete state calculations.
 Calculations of 
the $^{12}$C($p,n$) and $^{12}$C($n,p$) reactions which took binding
energy effects into account via the use of Woods-Saxon 
single-particle wave functions found 
a $\sim\! 30\%$  suppression of the predicted cross sections 
over the predictions using harmonic oscillator
wave functions.
In the present work we believe it is essential to work with Woods-Saxon radial
functions, but for comparison purposes we will give some results with
harmonic oscillator single-particle functions.

\section{Shell-model Calculations}
\label{s:shell}
\subsection{Model Space}
\label{ss:MS}

To investigate the effect of the approximations
inherent in any model calculations
of the inclusive neutrino cross sections on carbon we mount
a series of shell-model calculations, and 
present
calculations in four separate model spaces.
The spaces include excitations up to $4\hbar\omega$,
and we label these model spaces
(1) Closed-shell TDA, (2) Closed-shell RPA, 
(3) Closed-shell RPA +2p-2h, and (4) Unrestricted shell model. The labels
represent the model spaces:

1. {\em Closed-shell TDA}.  The $^{12}$C ground state is described as a
closed $0p_{3/2}$ shell and the $^{12}$N excitations as one 
particle-one hole (1p-1h) states:

\bea
\mid ^{12}\!C \rangle & = & \mid 0 \rangle ,
\nonumber \\
\mid ^{12}\!N \rangle & = & \mid (h^{-1},p)^{n \hbar \omega} J,T \rangle
~~~~~~~ n = 0,1,2,3,4
\label{TDA}
\eea

\noindent The 1p-1h states are labeled by their energy of excitation,
$n \hbar \omega$, in an oscillator model.  For example, states
$(p_{3/2}^{-1}, p_{1/2})$ are $0 \hbar \omega$ excitations,
$(s_{1/2}^{-1}, p_{1/2})$ and $(p_{3/2}^{-1}, sd)$ are
$1 \hbar \omega$ excitations, 
$(s_{1/2}^{-1}, sd)$ and $(p_{3/2}^{-1}, pf)$ are
$2 \hbar \omega$ excitations, and so on.
We present calculations up to $4 \hbar \omega$ excitation.
A shell-model calculation in this model space is equivalent to
the Tamm-Dancoff Approximation (TDA) for particle-hole excitations. 

2. {\em Closed-shell RPA}.  In this case, 2p-2h excitations are
included in the $^{12}$C ground-state wave function

\bea
\mid ^{12}\!C \rangle & = & \mid 0 \rangle + 
\mid (h_1^{-1},h_2^{-1}) J_1,T_1; (p_1,p_2) J_1,T_1 : 00 \rangle
\nonumber \\
\mid ^{12}\!N \rangle & = & \mid (h^{-1},p)^{n \hbar \omega} J,T \rangle
~~~~~~~ n = 0,1,2,3,4
\label{RPA}
\eea

\noindent Here $h_1,h_2$ span the hole orbitals, $0s_{1/2}$ and          
$0p_{3/2}$, while $p_1,p_2$ span particle orbitals
$0p_{1/2}$, 
$0d_{5/2}$, 
$1s_{1/2}$, $\ldots$.  The highest-energy orbital included in the
particle space matches the highest-energy 
orbital, $p$, in the 1p-1h basis of energy $n \hbar \omega$.
Although we have labeled this calculation, RPA, there are more 2p-2h 
states included in the $^{12}$C wave function here than are normally
present in an RPA calculation.  This is because the 2p-2h
correlations introduced in RPA are restricted to the type

\begin{displaymath}
\mid (h_1^{-1},p_1)^{n \hbar \omega } J,T;
(h_2^{-1},p_2)^{n \hbar \omega } J,T: 00 \rangle ,
\end{displaymath}

\noindent where the 2p-2h states are made up only from the coupling
of two 1p-1h states of spin, isospin $J,T$ that comprise the
basis states of $^{12}$N.  Not only is this 2p-2h basis smaller
than that in Eq.\ (\ref{RPA}) it also is not fully antisymmetrized.
Hole states $h_1$ and $h_2$, and particle states $p_1$ and $p_2$ are 
not antisymmetrized with respect to each other.  This shortcoming
is not present in the basis of Eq.\ (\ref{RPA}), where the
coupling order shown makes it easy to antisymmetrize the states.
Although, as we have just explained, a calculation in this model space
is more than just RPA, this case is the closest we have to the
calculations of Kolbe \etal \cite{Ko95}.

3. {\em Closed-shell RPA + 2p-2h}.  In this case, we add some 2p-2h
configurations to the 1p-1h basis states of $^{12}$N:

\bea
\mid ^{12}\!C \rangle & = & \mid 0 \rangle + 
\mid (h_1^{-1},h_2^{-1}) J_1,T_1; (p_1,p_2) J_1,T_1 : 00 \rangle
\nonumber \\
\mid ^{12}\!N \rangle & = & \mid (h^{-1},p)^{n \hbar \omega} J,T \rangle
+ \mid (p_{3/2}^{-1}, h^{-1})J_3,T_3;
(p_{1/2}, p)J_4,T_4: J,T \rangle
~~~~~~~ n = 0,1,2,3,4
\label{RPA22}
\eea

\noindent Note that among the 2p-2h configurations for $^{12}$N,
one of the holes is restricted to the $0p_{3/2}$ orbital and one
of the particles to the $0p_{1/2}$ orbital.  In this way,
we are moving further away from the assumption that the $^{12}$C
wave function is a closed $p_{3/2}$ shell.  Again the other hole orbital,
$h$, may
span the hole orbitals, $0s_{1/2}$ and          
$0p_{3/2}$, while the other particle orbit, $p$, spans orbitals
$0p_{1/2}$, 
$0d_{5/2}$, 
$1s_{1/2}$, $\ldots$ up to the same maximum characterizing the 1p-1h states.

4. {\em Fully Unrestricted Shell Model}. Finally, we move to an unrestricted 
shell-model calculation, with full mixing between all configurations:
\begin{eqnarray}
\mid ^{12}\!C \; g.s. >& = &\mid (p_{3/2}p_{1/2})^8: 00>  +
  \mid (p_{3/2}p_{1/2})^6 (1s0d)^2: 00> \nonumber \\
&   + &  \mid (p_{3/2}p_{1/2})^7 (1p0f): 00>  
+ \mid (0s)^{-1}(p_{3/2}p_{1/2})^8 (1s0d): 00>
\nonumber\\
& & \nonumber \\
\mid ^{12}\!N \;\;  \pi+ > &  = & \mid (p_{3/2}p_{1/2})^8: JT>  +
  \mid (p_{3/2}p_{1/2})^6 (1s0d)^2: JT> \nonumber\\
&  +  & \mid (p_{3/2}p_{1/2})^7 (1p0f): JT>  
+ \mid(0s)^{-1}(p_{3/2}p_{1/2})^8 (1s0d): JT> \nonumber \\
 & & \nonumber \\
\mid ^{12}\!N \;\; \pi- > & = & \mid (p_{3/2}p_{1/2})^7 (1s0d) : JT>
+ \mid (0s)^{-1} (p_{3/2}p_{1/2})^9 : JT> . 
\label{openRPA}
\end{eqnarray}
Spurious center-of-mass states were eliminated exactly by
making a transformation to an orthonormal basis for which
the expectation value of the centre-of-mass Hamiltonian is zero,
$< H_{cm} >= 0$. Our $2\hbar\omega$ shell model basis contains $\sim\! 2000$
states for each multipole, which is reduced by $\sim\! 25$ when transformed to
a non-spurious basis.
Because of computational limitations we are only able to take our
``unrestricted'' shell model calculations up to $2\hbar\omega$.
However, we assume an extrapolation of a $2\hbar\omega$ shell-model
calculation to
$4\hbar\omega$ scales in the same way as the closed-shell RPA $+$
2p-2h calculation.

\subsection{Effective Interactions}
\label{ss:effint}

Having selected the appropriate model spaces, we must specify the
operative effective interactions to use in these spaces.  We choose
interactions that have been fitted to reproduce experimental
spectra, namely:  For interactions among the $p$-shell orbitals,
we use the interaction of Cohen and Kurath \cite{CK65}, (8-16)2BME;
for interactions among the $s,d$-orbitals, we use the universal
$s,d$-interaction (USD) of Wildenthal \cite{USD}; for interactions
among the $p,f$-shell orbitals, we use the $G$-matrix of
Kuo and Brown \cite{KB68} as modified by Zuker \cite{PZ81} and
known as KB3.  All other two-body matrix elements, including
the cross-shell interactions, were calculated using the
Millener-Kurath \cite{MK75} parameterization in terms of central,
spin-orbit, and tensor interactions, whose radial forms are Yukawa
functions of various strengths and ranges in different spin-isospin
channels.  The parameters are determined in fits to spectra of
unnatural parity states in $p$-shell nuclei, and so are
particularly appropriate for the cross-shell interactions
between the $p$- and $s,d$-shell shells.  We, however,  use
this interaction for all cross-shell matrix elements required.

Finally, the single-particle energies must be specified.  For the
(8-16)2BME, USD and KB3 effective interactions these energies are given
as part of the fitted interaction, but are referenced to their
respective cores in their shell-model usage, namely
$^4$He, $^{16}$O and $^{40}$Ca.  In the present work, we wish
to make our reference core a closed $p_{3/2}$ shell at $^{12}$C,
so these single-particle energies are shifted according to the
formula

\be
\epsilon_j^A = \epsilon_j^B + \sum_{h,J,T}
\frac{(2J+1)(2T+1)}{2(2j+1)}
\langle (j,h) J,T \mid V \mid (j,h) J,T \rangle ,
\label{spe}
\ee

\noindent where  $\epsilon_j^A$ is the single-particle energy
of an orbital, $j$, relative to a core $A$,
$\epsilon_j^B$ 
relative to a core $B$, $A>B$, and the sum, $h$, is over all the
states occupied in $A$ but unoccupied in $B$.  Since only the
relative separation of single-particle energies is relevant in a shell-model
calculation we set the single-particle energy of the $0p_{1/2}$
orbital at its experimental value of $-4.95$ MeV.  For orbitals
above the $p,f$-shell, we have little guidance in choosing the
single-particle energies.  We use, therefore, a formula
given in Bohr-Mottelson \cite{BM75}

\be
\epsilon_{n,l,j} = const. +
v_{ls} \hbar \omega ({\bf l} . {\bf s} ) +
v_{ll} \hbar \omega \left ( {\bf l}^{2} - \langle {\bf l}^{2} \rangle_{N} 
\right ) ,
\label{vls}
\ee
\be
\langle {\bf l}^{2} \rangle_{N} = \frac{1}{2} N ( N + 3 ) ,
\label{ll}
\ee

\noindent where $N = 2 n + l$ is the principal quantum number for the
oscillator orbital, $n$ the number of radial nodes (excluding the
origin and infinity) and $l$ the orbital angular momentum quantum number.
We use the values $v_{ls} = -0.127$ and
$v_{ll} = -0.03$ from \cite{BM75}.  
The choice of $\hbar \omega$ to be used in this formula
should be one appropriate for nuclei where the
$(s,d,g)$- and $(p,f,h)$-orbitals are the valence orbitals, not
an $\hbar \omega$ appropriate for $^{12}$C.  We choose
$\hbar \omega = 7.2$ MeV.  The constant in Eq.\ (\ref{vls})
is again chosen so that when the formula is used for
the $0p_{1/2}$ orbital it reproduces the experimental
value, $-4.95$ MeV.
The values of the single-particle energies used are given in Table\
\ref{t:tab1}.

\section{Formalism}
\label{s:form}

We  apply the results of shell-model calculations just described to
the evaluation of a number of isovector weak and electromagnetic
transitions in the $A=12$ system.  We consider the inclusive
neutrino reactions

\be
\nu_{\mu} + ^{12}\!C \rightarrow X + \mu^{-},
\label{muDIF}
\ee
\be
\nu_e + ^{12}\!C \rightarrow X + e^{-},
\label{eDAR}
\ee

\noindent where $X$ is unobserved, the muon capture process

\be
\mu^{-}(1S) + ^{12}\!C \rightarrow X + \nu_{\mu} ,
\label{mucapt}
\ee

\noindent and photoabsorption

\be
\gamma + ^{12}\!C \rightarrow X ,
\label{photoabs}
\ee

\noindent
which proceeds only through the 
isovector E1 multipole. In addition, we take
guidance from earlier 
calculations\cite{djm1,djm2} for the 
($p,n$), ($n,p$) and ($e,e'$) reactions
to the isovector giant resonances of interest. Further we 
consider the exclusive cross section to the ground state
of $^{12}$N for the first three of the above reactions, as well as
the Gamow-Teller beta decay 
from $^{12}$N to the ground state of $^{12}$C.

The formalism for calculating these cross sections has been given by
O'Connell, Donnelly and Walecka \cite{ODW72}.  For the neutrino
absorption reactions, Eqs.\ (\ref{muDIF}) and (\ref{eDAR}), the
cross section is given by

\be
\sigma (E_{\nu}) = \frac{G^2}{2 \pi} \sum_f \int_{-1}^{+1}
d(\cos \theta ) \delta (E_i - E_f + E_{\nu} - E_e )
p_{\ell} E_{\ell} \mid \! M \! \mid^2,
\label{nusig}
\ee

\noindent where $G$ is the weak interaction coupling constant,
$G/(\hbar c )^3 = 1.16639 \times 10^{-5}$ GeV$^{-2}$,
$p_{\nu},E_{\nu}$ are the neutrino momentum and energy in the
lab system, $p_{\ell},E_{\ell}$ are the outgoing lepton momentum and energy,
$\cos \theta$ is the cosine of the angle between the electron
and neutrino directions, and $E_i-E_f$ is the mass difference
between the initial and final nuclei.  It is convenient to
define an energy and momentum transfer:  $q_0 = E_i-E_f = E_{\ell}-
E_{\nu}$ and ${\bf q} = {\bf p}_{\ell} - {\bf p}_{\nu}$.  There is
a sum over all final nuclear states, which in our calculations becomes
a sum over all the states available in a given shell-model
calculation. 

Lastly $\mid \! M \! \mid^2$ is given schematically

\be
\mid \! M \! \mid^2 = \sum_{\lambda =0}^{\infty} (
L_1 W_1^{(\lambda)} +
L_2 W_2^{(\lambda)} +
L_3 W_3^{(\lambda)} +
L_4 W_4^{(\lambda)} +
L_5 W_5^{(\lambda)} )
\label{m2}
\ee

\noindent where $L_1 \ldots L_5$ are the five lepton traces given in
Table II of Ref.\ \cite{ODW72} and are 
functions of $q_0$, ${\bf q}$,
and $\cos \theta$, while $W_1^{(\lambda)} \ldots W_5^{(\lambda)}$ 
are certain
combinations of squares of nuclear matrix elements.  The sum is over 
all multipolarities of operators, $\lambda$, satisfying angular
momentum condition $\Delta (J_i, J_f, \lambda)$, where $J_i$ and $J_f$
are the initial and final spins.

In general, there are seven transition operators, 
${\cal O}^{(\lambda)}(q{\bf x})$,
detailed in the tables of Donnelly and Haxton \cite{DH79}. 
The operators are functions of $q$, the magnitude of
the momentum transfer, $ q = \mid \! {\bf q} \! \mid$, and ${\bf x}$,
the position coordinate.  Reduced matrix elements of these operators
can be decomposed into two factors:

\be
\langle J_f \parallel \! {\cal O}^{(\lambda)}(q {\bf x}) \! \parallel J_i
\rangle = \sum_{j_{\alpha},j_{\beta}} \langle J_f \parallel \!
[ a_{j_{\alpha}}^{\dagger},a_{j_{\beta}} ]^{(\lambda)} \! \parallel 
J_i \rangle
\langle j_{\alpha} \parallel \! {\cal O}^{(\lambda)}(q {\bf x}) \!
\parallel j_{\beta} \rangle ,
\label{dbsp}
\ee

\noindent where the accompanying isospin quantum numbers have been
suppressed for economy of notation.  The first factor is the
expectation value of shell-model creation
and annihilation operators evaluated with the many-body shell-model
wave functions and is known as the one-body density matrix element (obdme).
The second factor, the single-particle matrix element (spme), is
independent of the many-body wave functions but depends on the
transition process in question through the momentum transfer, $q$.  This
factorization enables the shell-model calculations to be separated 
from the transition rate calculations.  Files of obdme prepared
in the former are transferred to the latter. The spme also
involve radial integrals of spherical Bessel functions, $j_L(qx)$,
or their derivatives with two single-particle nuclear radial
functions.  For the latter, we use Woods-Saxon functions.  
The nuclear matrix elements
are also multiplied by form factors, which are functions of $q$,
as given in Ref.\ \cite{ODW72}.

In the shell-model calculations,  the 
center-of-mass is treated on an equal footing with 
all other $3(A-1)$ coordinates, and which
causes  an erroneous center-of-mass contribution to the 
neutrino cross sections and muon capture rate.  
To  correct for this
we  multiply
the computed cross section by the 
Tassie-Barker function $\mid \! g_{\rm cm} (q) \!
\mid^2$, where

\be
g_{\rm cm}(q) = e^{y/A}
\label{cmcorr}
\ee

\noindent with $y = (bq/2)^2$, $q$ the momentum transfer, $b$ the 
oscillator length parameter and $A$ the nuclear mass number.  
For $y \sim 1$
and $A = 12$, this correction provides a 20\% increase in the
computed cross section.\footnote{This correction has been
derived with the use of oscillator functions.  We apply the same 
correction even when using Woods-Saxon radial functions.}

In the expression for the neutrino absorption cross-section,
Eq.\ (\ref{nusig}), one further correction has to be applied
that represents the distortion of the outgoing lepton in the
Coulomb field of the daughter nucleus.  We are guided by
the work of Engel \cite{En98} in this matter.  The energy of
the emerging lepton, while under the influence of a constant
electrostatic potential within the nucleus, is effectively shifted:

\bea
E_{\ell ,{\rm eff}} & = & E_{\ell} - V(0) 
\nonumber  \\ 
p_{\ell ,{\rm eff}} & = & \left ( E_{\ell ,{\rm eff}}^2
- m^2 \right )^{1/2}
\nonumber  \\ 
V(0) & = & - \frac{3}{2} \frac{Z \alpha}{R}
\label{Vee0}
\eea

\noindent where $E_{\ell}$, $p_{\ell}$ and $m$ are the energy,
momentum and mass of the emerging lepton, $Z$ the charge number for
the daughter nucleus (with a plus sign for particle leptons, and
a minus sign for antiparticle leptons), $R$ the nuclear radius
and $\alpha$ the fine-structure constant.  If $p_{\ell} R$ is
less than $\sim 0.5$, then the lepton wave function is
primarily $s$-state, and the usual Fermi function, $F(Z,E_{\ell})$
is used as a multiplicative correction factor in Eq.\ (\ref{nusig}).
For DAR cross sections, $p_{\ell} R$ is always less than 0.5.
For DIF cross sections, the emerging muons have
$p_{\ell} R > 0.5 $ for most of the energy range.  Then we follow
Engel's suggestion \cite{En98} that the cross-section be multiplied 
by a factor
$p_{\ell ,{\rm eff}} 
E_{\ell ,{\rm eff}} / (
p_{\ell} 
E_{\ell} )$
and the momentum transfer $q = |{\bf q}|$ used in the spherical
Bessel functions in the radial integrals be shortened by using
${\bf q}_{{\rm eff}} = {\bf p}_{\ell ,{\rm eff}} - {\bf p}_{\nu}$.
The multiplicative factor increases the DIF cross section
by about 16\% while the shortening of $q$ in the Bessel functions
reduces this enhancement to about 11\%.

In the results to be given in the next section, for the neutrino
cross sections Eqs.\ (\ref{muDIF}) and (\ref{eDAR}), we will give
the flux-averaged cross-sections defined as

\be
\overline{\sigma} = \int d E_{\nu} \Lambda (E_{\nu}) \sigma (E_{\nu}) /
\int d E_{\nu} \Lambda (E_{\nu}) ,
\label{fluxav}
\ee

\noindent where $\Lambda (E_{\nu})$ is the incident neutrino flux.
For the liquid scintillator neutrino detector (LSND) 
experiment \cite{At97},
we use their most recently determined muon-neutrino
flux \cite{LSNDflux} from pion decays in flight. For the experiments
by the KARMEN Collaboration \cite{Bo94},
the electron-neutrino beam comes from muon decays at rest and
the flux is given by the Michel spectrum.

The muon capture rate, Eq.\ (\ref{mucapt}), is given by \cite{ODW72}

\be
\Lambda_c = \frac{G^2 E_{\nu}^2}{2 \pi} \mid \! \phi_{1S} \!
\mid_{\rm av}^2 \left ( 1 + \frac{E_{\nu}}{m_T} \right )^{-1}
\mid \! M \! \mid^2 ,
\label{murate}
\ee

\noindent where the outgoing neutrino energy is $E_{\nu} = m_{\ell} -
\epsilon_b +E_i - E_f$, with $m_{\ell}$ the muon reduced mass, 
$\epsilon_b$ the muon K-shell binding energy and $E_i - E_f = q_0$
is the mass difference between the initial and final nuclear states.
The factor $(1 + E_{\nu}/m_T)^{-1}$ is a recoil correction, with
$m_T$ the target nucleus mass.
Further, $\phi_{1S}$ is the muon K-shell bound state wave function
evaluated at the nucleus 

\be
\mid \! \phi_{1S} \mid_{\rm av}^2 = \frac{R}{\pi} (Z \alpha m_{\ell} )^3
\label{Kwfn}
\ee

\noindent with 
$R$ a reduction factor for the finite size
of the nuclear charge distribution, $R = 0.86$ for $^{12}$C.
The total inclusive capture rate is the sum of Eq.\ (\ref{murate}),
summed over all final states contained in the shell-model calculation.

Finally, the photoabsorption cross section, Eq.\ (\ref{photoabs}), is
given by \cite{BM75} 

\be 
\sigma (E_{\gamma}) = \frac{16 \pi^3}{9} \alpha E_{\gamma}
B(E1; i \rightarrow f) \delta ( E_{\gamma} -E_f + E_i )
\label{sigphoto}
\ee

\noindent where $E_{\gamma}$ is the photon energy and $B(E1; i \rightarrow
f )$ the reduced transition probability for $E1$ photon absorption.
Again the total absorption cross section, $\sigma_{\rm tot}$, is
given by 
the sum of Eq.\ (\ref{sigphoto}),
summed over all final states contained in the shell-model calculation.
In the different calculations, the total absorption cross section
reaches a saturation value at a photon energy $E_{\gamma}$ of
the order of 50 MeV, slightly less in $1 \hbar \omega$ calculations,
slightly more in $(1 + 3) \hbar \omega$.  Note, only the $E1$ multipole
is retained in the calculations.  The experimental total absorption
cross section\cite{E1expt}, however, shows no sign of saturating
and keeps growing with increasing $E_{\gamma}$.  This is because 
higher multipoles are contributing.  In quoting an experimental
value, we have cut off the contributions at $E_{\gamma} = 50 \pm 5$
MeV for a total cross section of $\sigma_{\rm tot} = 21 \pm 1$ fm$^2$.

\section{Results}
\label{s:rests}
\subsection{Inclusive Cross-sections}
\label{ss:ICs}

We present the results of our calculations in Tables \ref{t:tab2} to 
\ref{t:tab5}. In most of our tables the  tabulated results 
are for Woods-Saxon radial wave functions. In Table \ref{t:tab3}
we list the results
for three sets of radial wave functions,  Woods-Saxon, and 
harmonic oscillators
with $b=1.82$ fm and $b=1.64$ fm.
For each of the four model spaces, we give the
results as a function of the size of the particle-hole space, increasing 
from $0 \hbar \omega$ excitations up to $4 \hbar \omega$.
In Table \ref{t:tab2} are the inclusive cross sections for
$(\nu_{\mu},\mu^{-})$-scattering,
$(\nu_e,e^{-})$-scattering,
muon capture, and photoabsorption on $^{12}$C.  The calculation in the
TDA approximation overestimates the cross sections in all cases, while
the two variants of RPA reduce these cross sections by about a
factor of two and are in closer agreement with experiment.
For the
$(\nu_{\mu},\mu^{-})$ cross section,
the calculations  to $4 \hbar \omega$ in the closed-shell RPA,
closed-shell RPA + 2p-2h, and unrestricted shell model
yield 18.2, 16.7 and
$13.8 \times 10^{-40}$ cm$^2$, the latter value being an
extrapolation
from a $2 \hbar \omega$ calculation.  This latter value,
also, is within range of the experimental value 
of $(12.4 \pm 1.8) \times 10^{-40}$ cm$^2$ \cite{At97}. 
This is our principal result.
In ref\cite{Ko95},
 with the $p_{3/2}$ subshell taken as a closed shell,
Kolbe \etal ~obtain a result $19.8(5) \times 10^{-40}$ cm$^2$,
where the uncertainty represents a spread between different choices of the
effective interaction.  Recently \cite{Ko99}, this work has been
extended to include a partial occupation of the $p_{1/2}$
subshell obtaining a result some two units smaller of
$17.6(2) \times 10^{-40}$ cm$^2$.  The present calculation
 reduces this
four  units further with the use 
of fully unrestricted shell-model configurations and Woods-Saxon
radial wave functions.

However, in obtaining an improved result for the DIF cross section,
we must check that the same calculation still satisfactorily reproduces
other relevant data.  In Table \ref{t:tab2} we therefore give
results for $(\nu_e,e^{-})$ DAR cross sections, $\mu$-capture
rates and photoabsorption cross sections.  For
$(\nu_e,e^{-})$ neutrino absorption, we calculate in 
closed-shell RPA, closed-shell RPA $+$ 2p-2h, and
unrestricted shell-model cross sections of 21.9, 20.4 and 12.5 $\times
10^{-42}$ cm$^2$ respectively compared with the
KARMEN result of $(14.0 \pm 1.2) \times 10^{-42}$ cm$^2$
and the LSND result of $(14.8 \pm 1.3) \times 10^{-42}$ cm$^2$.
Again the unrestricted shell-model calculation gives a big reduction to
the neutrino absorption cross sections resulting, in this case,
with a value that is on the low side compared to
experiment.  However, it is only outside the experimental range
by one standard deviation.  By comparison, the continuum 
RPA calculation of Kolbe \etal \cite{Ko99} obtains a result
$14.4 \times 10^{-42}$ cm$^2$ exactly in the right range.
Thus it would seem that in obtaining a smaller DIF cross
section as required, we are simultaneously underpredicting the
DAR cross section.

For $\mu$-capture and photoabsorption, our results in
Table \ref{t:tab2} indicate the unrestricted shell-model 
calculation slightly
overpredicting the experimental result, but only by about
10\%.  In summary, then, our improved result for the DIF
cross section comes with some modest deterioration
in the DAR, $\mu$-capture and photoabsorption results.

\subsection{The Radial Wave Functions}
\label{ss:RWF}

In discrete-state shell-model calculations with effective
interactions, the choice of radial wave functions to use in evaluating
transition matrix elements remains unspecified.  There is
therefore some freedom in making this choice.  In Table \ref{t:tab3}
we give results 
for three choices of radial functions 
for the inclusive cross sections, but in
this case explicitly exclude the ground-state contribution.

A description of the rms charge radius of $^{12}$C using 
harmonic oscillator single-particle wave functions requires
an oscillator parameter of $b=1.64$ fm.  
However, to obtain reasonably good fits to the 
shape of the electron-scattering form factors in $^{12}$C
considerably different values of $b$ are often needed. 
An analysis\cite{comfort,djm1,djm2} of
the ($e,e'$) form factors and ($p,p'$) inelastic scattering data suggests
the need for state dependent oscillator parameters 
in the broad range $1.64 - 1.94$ fm.
The problem arises as a result of the inability to 
describe loosely bound or unbound states of $^{12}$N 
using the same oscillator parameter as is needed to 
describe the deeply bound nucleons in the $^{12}$C ground state.

In a systematic study of the effect of more realistic 
single-particle wave function on the
$^{12}$C$(p,n)$ reaction Millener \etal  \cite{djm1,djm2} 
concluded that the use of Woods-Saxon wave functions
reduces the predicted cross sections by about 30$\%$ over 
the harmonic oscillator predictions.
 Ohnuma \etal \cite{ohnuma}
drew a similar conclusion. The large suppression of the 
cross section with the use of Woods-Saxon single particle 
wave functions is because of 
the large mismatch between the radial wave functions describing
the $^{12}$C$_{g.s.}$ versus the $^{12}$N states.
An effect of similar magnitude was found for 
the $^{12}$C($e,e'$) form factors.

The cross sections obtained using Woods-Saxon wave functions 
have been found \cite{comfort,haxton}
to be quite similar to those obtained when the  
harmonic oscillator parameter is chosen to fit best the
($e,e'$) form factor. Brady \etal \cite{djm1} obtained a good fit
 to the ($e,e'$) form factor for 
the $^{12}$C$_{g.s.}\rightarrow ^{12}$C(15.11 MeV)
using an oscillator parameter $b=1.82$ fm.
To examine the sensitivity of the predicted neutrino cross sections to 
the assumed oscillator parameter we recalculated 
the cross sections  using a smaller
value of $b$, and the results are tabulated  in Table \ref{t:tab3}.
Relative to the $b=1.82$ fm calculation the 
inclusive ($\nu_\mu,\mu^-$) DIF cross section
increased only slightly for $b = 1.64$ fm,
while the $(\nu_e,e^-$) DAR cross section decreased by 17$\%$
and the $\mu$-capture rate decreased by 14$\%$. The larger 
value of $b$ in all cases gave
predictions closer in agreement with experiment.

The difference in the response of the three  
reactions to the larger 
oscillator parameter reflects the different average momentum 
transfer involved in each, DIF ($q\sim1$ fm$^{-1}$),
DAR ($q\sim 0.2$ fm$^{-1}$), and $\mu$-capture ($q\sim 0.5$ fm$^{-1}$).
Changing the oscillator parameter changes the shape of the 
axial-vector and vector form factors
entering the neutrino cross sections.
For momentum transfers below the first maximum of the form 
factor the neutrino
cross section increases as $b$ increases. In contrast, momentum 
transfers beyond
the first peak in the form factor result in the 
predicted cross section being decreased as $b$ increases.

We next consider the use of Woods-Saxon radial wave functions.
The mismatch between the initial and final 
radial wave function, resulting
from binding energy changes, cannot be accounted for 
within a harmonic oscillator basis.
It is impossible to treat the binding energy of the final nucleon 
rigorously within a discrete shell model, but 
an estimate of the effect of
binding energies on the predicted cross sections can be obtained using
Woods-Saxon wave functions that are only just bound. 
The Woods-Saxon
result reduces the  predicted inclusive DIF ($\nu_\mu,\mu^-)$
cross section by about six
units, 
putting the result within the experimental
range.   However the predicted DAR $(\nu_e,e^-)$
 cross-section is similarly reduced
by 2.8 units and is too small compared with experiment.
The calculated  $\mu$-capture rate increases  and slightly 
overpredicts experiment.
 
The use  of  Woods-Saxon functions causes some radial integrals to become
non-zero, where
they are identically zero for oscillator wave functions.
This is seen easily for the E1 
$\gamma$-ray  transition matrix elements. 
With oscillators, the predicted B(E1) strength in $(0+1+2+3)\hbar\omega$
 is slightly
less than in a $(0+1)\hbar\omega$ calculation. But with the use
of Woods-Saxon wave functions  the 
$(0+1+2+3)\hbar\omega$ calculation is about 40\% larger. 
This is due to the non-vanishing
integral $\int R(n,l) r R(n',l') r^2 dr$, 
when radial functions, $R(n,l)$ and $R(n',l')$, represent
single-particle states differing in energy
by $3\hbar\omega$.  The same effect is noticeable in the 
calculated  $\mu$-capture rate  and to some
extent in neutrino scattering cross sections.

\subsection{The Exclusive Cross Section to the $^{12}$N$_{g.s.}$}
\label{ss:ECS}

Next, we examine in detail the exclusive cross section to the $^{12}$N
ground state.
The results are in Table \ref{t:tab4}.  The first row
of the TDA calculation labeled $0 \hbar \omega$ is the
single-particle $0^{+} \rightarrow (p_{3/2}^{-1},p_{1/2}) 1^{+}$
transition rate, while the first row of the unrestricted shell-model
calculations represents the complete $p$-shell mixing as
given by Cohen-Kurath (8-16)2BME wave functions.  We note that the
cross sections in the configuration-mixed calculations are
a factor four to five smaller than the single-particle estimate.
The closed-shell RPA variants are only able to obtain about a
factor of two of these reductions.  Increasing the particle-hole 
space to include $2 \hbar \omega$ and $4 \hbar \omega$ excitations
only has an impact in the $10 - 20\%$ range for
neutrino absorption,
muon capture and beta decay. 

In our complete unrestricted $(0+2)\hbar\omega$ calculations 
we paid particular 
attention to
handling the strong $\Delta\hbar\omega=2$
 interaction in a consistent way.
When  empirical or realistic Hamiltonians which give
a reasonable description of $0\hbar\omega$ and $1\hbar\omega$
spectra are used in full (0+2)$\hbar\omega$ spaces, the $\Delta\hbar\omega=2$ 
interaction pushes the ground
state down by several MeV. The resulting spectrum is in poor agreement with 
experiment.
With no restrictions placed on our (0+1+2)$\hbar\omega$
calculations we found that the giant monopole resonance
was predicted to lie too low in the spectrum and large percentages of $1p1h$ 
excitations were predicted
in the ground state wave function.
To correct for this pathological behavior we adjusted diagonal matrix elements 
of the Hamiltonian so as to 
restore the energy of the giant monopole resonance to lie above 20 MeV, as 
expected by self-consistent
RPA calculation using density-dependent interactions.
In addition we set the strong $\langle 1p1h \; 2\hbar\omega|H|0\hbar\omega 
\rangle$ matrix elements, that transform as
$(\lambda,\mu)=(2,0)$ under SU(3), to zero.
The resulting $^{12}$C ground-state wave function yields a predicted strength 
for
the GT transition to the  $^{12}$N$_{g.s.}$ that is in 
reasonable agreement with 
experiment, Table \ref{t:tab4}.

\subsection{The Multipole Decomposition}
\label{ss:MD}

Finally, in Table \ref{t:tab5} we give the breakdown in the contribution
to the cross sections from each multipole, for oscillator wave functions
using $b=1.82$ fm.  
The most striking result
is that 80\% of the
$(\nu_e,e^{-})$ inclusive cross sections
comes from the ground-state $1^{+}$ multipole, with the remaining
20\% coming from the excited state $1^{+}$, $1^{-}$, and $2^{-}$
multipoles.  Muon capture also is dominated by the ground-state $1^{+}$ 
multipole, being, in this case, 65\% of the inclusive rate.
The remaining 35\% is dominated by the $1^{-}$ multipole with
lesser contributions from the $2^{-}$, $0^{-}$, excited $1^{+}$,
and $2^{+}$ multipoles.  By contrast, the
$(\nu_{\mu},\mu^{-})$ inclusive cross section gets only a 10\%
contribution from the ground-state $1^{+}$ multipole, while the
remaining 90\% is spread over many multipoles,
$1^{-}$ to $4^{-}$ and $1^{+}$ to $3^{+}$ dominating.

\section{Conclusions}
\label{s:Conc}

We have examined the $(\nu_\mu,\mu^-$), $(\nu_e,e^-$), and $\mu$-capture
reactions on $^{12}$C within a shell-model calculations 
that included up to 
$4\hbar\omega$ of excitation. Calculations were carried 
out for four separate model spaces,
 for three sets of radial wave functions.
 Closest agreement 
between theory and experiment 
was obtained for the most
sophisticated model spaces,  
which 
included both RPA-type correlations and an open
$p$-shell $^{12}$C ground state. 

Previous analyses of the ($e,e'$), ($p,n$) and ($n,p$) reactions
have shown that when the large difference in binding 
energies between the $^{12}$C ground state and the $^{12}$N
final states in taken into account,
the predicted cross sections are reduced by about $30\%$ 
relative to harmonic oscillator shell-model 
results.
To examine the effect of more realistic single-particle wave functions
we compared the predictions using two oscillator parameters, 
$b=1.64$ fm and $b=1.82$ fm and Woods-Saxon wave functions. 
The lower value of $b$ is suggested by the $^{12}$C ground-state 
charge radius, while the larger value comes
from a fit to the shape of the transverse form factor 
for the $^{12}$C($0\rightarrow 15.11; 1^+$) transition.
Increasing $b$ increases the predicted cross section 
for all low momentum transfer reactions.
The photoabsorption 
cross section increases with $b^2$, and the 
DAR $(\nu_e,e^-)$ cross-section increased by a smaller but
similar amount.
For higher momentum transfer processes ($q \sim 1$ fm$^{-1}$), 
such as the DIF neutrino reaction and the 
($p,n$) reaction at intermediate energies, the predicted cross section 
decreases with increasing
$b$.
Of course, binding energy differences result in a mismatch 
between the initial and final radial wave functions 
which  cannot be mimicked by a change in the 
oscillator parameter, and fits to the 
($e,e'$) form factors for states in $^{12}$C within an oscillator basis 
find a need for state-dependent oscillator parameters in 
the range $1.64-1.94$ fm. Incorporating binding energy 
effects through the use of Woods-Saxon wave function 
has a large effect on the predicted cross sections
for all reactions.
The predicted inclusive DIF $(\nu_\mu,\mu^-)$ cross section is 
in  agreement with the experimental range. However, 
the Woods-Saxon calculation
underpredicts the DAR $(\nu_e,e^-)$ cross section and overpredicts
the $\mu$-capture rate.

Given the model dependence noted in the present set of calculations,
it is difficult to determine
the expected DIF cross section to high accuracy using 
constraints from other observables.
For example, the ratio of the predicted DAR to DIF cross sections 
changed by about 20$\%$ for the
two different values of oscillator parameter considered.
Furthermore, we were unable to include  a fully realistic treatment 
of the continuum nature of
states of $^{12}$N. 
However the success of the Woods-Saxon calculation for the DIF
cross section while only failing to obtain the experimental
result for the DAR cross section, $\mu$-capture rate and
photoabsorption cross section by one to two standard
deviations indicates the importance of a continuum treatment for 
the final states, coupled with an open-shell description
of the nuclear configurations.

\begin{table}
\begin{center}
\caption{ Spin structure of the $^{12}$C ground state
\label{t:tab0}}
\begin{tabular}{lllll}
& & & & \\[-3mm]
Model &
S = 0 &
S = 1 &
S = 2 &
$\% [\widetilde{444}]  $ \\[3mm]
\tableline
& & & & \\[-3mm]
Three $\alpha-$particles & 100$\%$ & 0 & 0 & 100$\%$\\[3mm]
Cohen-Kurath Interaction &
81$\%$ &
18$\%$ &
$< 2 \% $ &
78$\%$\\[3mm]
$ (p_{3/2})^8$ &
 16$\%$&
 59$\%$&
25$\%$& 6$\%$ \\
\end{tabular}
\vspace{-5mm}
\end{center}

\vspace*{.6cm }
\end{table}

\begin{table}
\begin{center}
\caption{Single-particle energies.
\label{t:tab1}}
\begin{tabular}{ccccccccccc}
& & & & & & & & & & \\[-3mm]
$0s_{1/2}$ &
$0p_{3/2}$ &
$0p_{1/2}$ &
$0d_{5/2}$ &
$1s_{1/2}$ &
$0d_{3/2}$ &
$0f_{7/2}$ &
$1p_{3/2}$ &
$0f_{5/2}$ &
$1p_{1/2}$ & \\
\tableline
& & & & & & & & & & \\[-3mm]
$-$34.40 & $-$12.26 & $-$4.95 & 5.98 & 3.70 & 8.97 & 10.16 &
8.78 & 9.39 & 8.39 & \\
\hline \\
& & & & & & & & & & \\[-3mm]
$0g_{9/2}$ &
$1d_{5/2}$ &
$2s_{1/2}$ &
$0g_{7/2}$ &
$1d_{3/2}$ &
$0h_{11/2}$ &
$1f_{7/2}$ &
$2p_{3/2}$ &
$0h_{9/2}$ &
$1f_{5/2}$ &
$2p_{1/2}$ \\
\hline \\
& & & & & & & & & & \\[-8mm]
13.87 & 18.09 & 20.46 & 18.28 & 20.54 & 
20.17 & 25.32 & 28.61 & 25.56 & 28.75 & 30.08 \\ 
\end{tabular}
\end{center}
\end{table}

\begin{table}
\begin{center}
\caption{Inclusive cross sections involving a continuum of $^{12}$N
states, including the ground state. Woods-Saxon radial functions are used,
with their asymptotic forms matched to the experimental separation
energies for bound states, and matched to a binding energy of 0.01 MeV
for unbound states.
\label{t:tab2}}
\begin{tabular}{ccccc}
& & & & \\[-3mm]
& $(\nu_{\mu},\mu^{-})$ DIF &
$(\nu_e,e^{-})$ DAR &
$\mu$-capture  &
photoabsorption  \\
& $\overline{\sigma} \times 10^{-40}$ cm$^2$ &
$\overline{\sigma} \times 10^{-42}$ cm$^2$ &
$\Lambda_c \times 10^3$ s$^{-1}$ &
$\sigma_{\rm tot} \times 10^{-26}$ cm$^2$ \\
\tableline
& & & & \\[-3mm]
Closed-shell TDA & & & & \\
$0 \hbar \omega$ & ~3.8 & 34.6 & 31.0 & \\
$(0 + 1) \hbar \omega $ & 12.8 & 39.0 & 62.9 & 22.7 \\
$(0 + 1 + 2) \hbar \omega $ & 26.1 & 40.0 & 67.7 &  \\
$(0 + 1 + 2 + 3) \hbar \omega $ & 31.0 & 40.7 & 74.0 & 29.5 \\
$(0 + 1 + 2 + 3 + 4) \hbar \omega $ & 33.7 & 40.6 & 75.9 & \\[3mm]
Closed-shell RPA & & & & \\
$0 \hbar \omega$ & ~1.7 & 13.3 & 11.7 & \\
$(0 + 1) \hbar \omega $ & ~6.9  & 15.8 & 35.1 & 14.4 \\
$(0 + 1 + 2) \hbar \omega $ & 13.2 & 20.9 & 41.4 &  \\
$(0 + 1 + 2 + 3) \hbar \omega $ & 16.2 & 21.4 & 43.5 & 18.3 \\
$(0 + 1 + 2 + 3 + 4) \hbar \omega $ & 18.2 & 21.9 & 45.4 & \\[3mm]
Closed-shell RPA + 2p-2h & & & & \\
$0 \hbar \omega$ & ~1.9 & 14.6 & 13.4 & \\
$(0 + 1) \hbar \omega $ & ~7.1 & 17.0 & 36.2 & 17.9 \\
$(0 + 1 + 2) \hbar \omega $ & 13.6 & 21.1 & 42.5 &  \\
$(0 + 1 + 2 + 3) \hbar \omega $ & 16.1 & 21.3 & 43.5 & 21.6 \\
$(0 + 1 + 2 + 3 + 4) \hbar \omega $ & 16.7 & 20.4 & 44.1 & \\[3mm]
Unrestricted shell model & & & & \\
$0 \hbar \omega$ & ~1.3 & ~7.1 & ~6.4 & \\
$(0 + 1) \hbar \omega $ & ~7.6 & 10.5 & 31.6 & 19.6 \\
$(0 + 1 + 2) \hbar \omega $ & 11.1 & 12.1 & 39.5 &  \\
$(0 + 1 + 2 + 3) \hbar \omega $ & (13.2)\tablenotemark[1] &  
(12.3)\tablenotemark[1] &  
(40.6)\tablenotemark[1] &  
(23.6)\tablenotemark[1] \\ 
$(0 + 1 + 2 + 3 + 4) \hbar \omega $ & (13.8)\tablenotemark[1] &  
(12.5)\tablenotemark[1] &  
(42.2)\tablenotemark[1] & \\[3mm]  
CRPA\cite{Ko99} & 17.6(2) & 14.4(1) & 38.0(7) & \\[3mm] 
& & $14.0(12)$ \cite{Bo94} & & \\
Expt. & $12.4(18)$ \cite{At97} & $14.8(13)$ \cite{Im98,At97}
& $38.9(9)$ \cite{Su87,Gi81} & 21(1)\cite{E1expt} \\
\end{tabular}
\vspace{-5mm}
\tablenotetext[1]{Estimate}
\end{center}
\end{table}

\begin{table}
\begin{center}
\caption{Inclusive cross sections involving a continuum of $^{12}$N
states, but excluding the ground state, for various choices of
radial wavefunctions: oscillator functions with length parameter
$b = 1.64$ fm or 1.82 fm, and Woods-Saxon functions.
\label{t:tab3}}
\begin{tabular}{ccccc}
& & & & \\[-3mm]
& $(\nu_{\mu},\mu^{-})$ DIF &
$(\nu_e,e^{-})$ DAR &
$\mu$-capture  &
photoabsorption  \\
& $\overline{\sigma} \times 10^{-40}$ cm$^2$ &
$\overline{\sigma} \times 10^{-42}$ cm$^2$ &
$\Lambda_c \times 10^3$ s$^{-1}$ &
$\sigma_{\rm tot} \times 10^{-26}$ cm$^2$ \\
\tableline
& & & & \\[-3mm]
Oscillator functions ($b = 1.64$ fm) & & & & \\
Closed-shell RPA + 2p-2h  & & & & \\
$0 \hbar \omega$ & ~1.5 & ~1.1 & ~1.5 & \\
$(0 + 1) \hbar \omega $ & ~9.4 & ~4.9 & 22.6 & 17.9 \\
$(0 + 1 + 2) \hbar \omega $ & 14.2 & ~5.3 & 25.9 &  \\
$(0 + 1 + 2 + 3) \hbar \omega $ & 16.5 & ~5.1 & 24.6 & 16.8 \\
$(0 + 1 + 2 + 3 + 4) \hbar \omega $ & 17.1 & ~5.5 & 25.0 & \\[3mm]
Unrestricted shell model  & & & & \\
$0 \hbar \omega$ & ~1.3 & ~0.3 & ~0.7 & \\
$(0 + 1) \hbar \omega $ & 11.0 & ~5.6 & 27.1 & 22.3 \\
$(0 + 1 + 2) \hbar \omega $ & 15.9 & ~5.7 & 28.9 &  \\
$(0 + 1 + 2 + 3) \hbar \omega $ & (18.5)\tablenotemark[1] &  
(~5.5)\tablenotemark[1] &  
(27.5)\tablenotemark[1] &  
(20.9)\tablenotemark[1] \\ 
$(0 + 1 + 2 + 3 + 4) \hbar \omega $ & (19.1)\tablenotemark[1] &  
(~5.9)\tablenotemark[1] &  
(27.9)\tablenotemark[1] & \\[3mm]  
\tableline
& & & & \\[-3mm]
Oscillator functions ($b = 1.82$ fm) & & & & \\
Closed-shell RPA + 2p-2h  & & & & \\
$0 \hbar \omega$ & ~1.6 & ~1.1 & ~1.6 & \\
$(0 + 1) \hbar \omega $ & ~8.9 & ~5.6 & 25.5 & 22.1 \\
$(0 + 1 + 2) \hbar \omega $ & 13.6 & ~5.9 & 29.5 &  \\
$(0 + 1 + 2 + 3) \hbar \omega $ & 16.3 & ~5.7 & 27.9 & 20.7 \\
$(0 + 1 + 2 + 3 + 4) \hbar \omega $ & 17.2 & ~6.1 & 28.3 & \\[3mm]
Unrestricted shell model  & & & & \\
$0 \hbar \omega$ & ~1.3 & ~0.3 & ~0.9 & \\
$(0 + 1) \hbar \omega $ & 10.3 & ~6.6 & 30.8 & 27.5 \\
$(0 + 1 + 2) \hbar \omega $ & 15.1 & ~6.6 & 33.0 &  \\
$(0 + 1 + 2 + 3) \hbar \omega $ & (18.1)\tablenotemark[1] &  
(~6.4)\tablenotemark[1] &  
(31.2)\tablenotemark[1] & 
(25.7)\tablenotemark[1] \\ 
$(0 + 1 + 2 + 3 + 4) \hbar \omega $ & (19.1)\tablenotemark[1] &  
(~6.9)\tablenotemark[1] &  
(31.7)\tablenotemark[1] & \\[3mm]  
\tableline
& & & & \\[-3mm]
Woods-Saxon functions & & & & \\
Closed-shell RPA + 2p-2h  & & & & \\
$0 \hbar \omega$ & ~0.8 & ~0.6 & ~1.3 & \\
$(0 + 1) \hbar \omega $ & ~6.0 & ~3.0 & 24.1 & 17.9 \\
$(0 + 1 + 2) \hbar \omega $ & 12.4 & ~3.1 & 27.7 &  \\
$(0 + 1 + 2 + 3) \hbar \omega $ & 14.9 & ~3.3 & 28.6 & 21.6 \\
$(0 + 1 + 2 + 3 + 4) \hbar \omega $ & 15.6 & ~3.5 & 29.9 & \\[3mm]
Unrestricted shell-model  & & & & \\
$0 \hbar \omega$ & ~0.7 & ~0.2 & ~0.8 & \\
$(0 + 1) \hbar \omega $ & ~7.0 & ~3.6 & 26.0 & 19.6 \\
$(0 + 1 + 2) \hbar \omega $ & 10.5 & ~3.7 & 32.9 &  \\
$(0 + 1 + 2 + 3) \hbar \omega $ & (12.6)\tablenotemark[1] &  
(~3.8)\tablenotemark[1] &  
(34.0)\tablenotemark[1] &  
(23.6)\tablenotemark[1] \\ 
$(0 + 1 + 2 + 3 + 4) \hbar \omega $ & (13.2)\tablenotemark[1] &  
(~4.1)\tablenotemark[1] &  
(35.6)\tablenotemark[1] & \\[3mm]  
CRPA\cite{Ko99} & 16.9(2) & ~5.5(1) & 32.0(7) & \\[3mm] 

& & $5.1(8)$ \cite{Bo94} & & \\
Expt. & $11.7(18)$ \cite{At97} & $5.7(8)$ \cite{Im98}
& $32.8(8)$ \cite{Su87} & 21(1)\cite{E1expt} \\
\end{tabular}
\vspace{-5mm}
\tablenotetext[1]{Estimate}
\end{center}
\end{table}

\begin{table}
\begin{center}
\caption{Exclusive cross sections involving the ground state of $^{12}$N
only.
Woods-Saxon radial functions are used, whose asymptotic forms are matched
to the experimental ground-state separation energies.
\label{t:tab4}}
\begin{tabular}{ccccc}
& & & & \\[-3mm]
& $(\nu_{\mu},\mu^{-})$ DIF &
$(\nu_e,e^{-})$ DAR &
$\mu$-capture  &
$\beta$-decay  \\
& $\overline{\sigma} \times 10^{-40}$ cm$^2$ &
$\overline{\sigma} \times 10^{-42}$ cm$^2$ &
$\Lambda_c \times 10^3$ s$^{-1}$ &
$B(GT; 1^{+} \rightarrow 0^{+} )$ \\
\tableline
& & & & \\[-3mm]
Closed-shell TDA & & & & \\
$0 \hbar \omega$ & ~2.74 & 34.5 & 30.0 & 2.26 \\
$(0 + 1 + 2) \hbar \omega $ & ~2.41 & 35.4 & 30.7 & 2.45 \\
$(0 + 1 + 2 + 3 + 4) \hbar \omega $ & ~2.42 & 35.2 & 30.6 & 2.44 \\[3mm]
Closed-shell RPA & & & & \\
$0 \hbar \omega$ & ~1.07 & 13.2 & 11.1 & 0.87 \\
$(0 + 1 + 2) \hbar \omega $ & ~1.19 & 18.2 & 15.1 & 1.26 \\
$(0 + 1 + 2 + 3 + 4) \hbar \omega $ & ~1.24 & 18.8 & 15.6 & 1.30 \\[3mm]
Closed-shell RPA + 2p-2h & & & & \\
$0 \hbar \omega$ & ~1.17 & 14.0 & 12.1 & 0.91 \\
$(0 + 1 + 2) \hbar \omega $ & ~1.16 & 18.0 & 14.9 & 1.23 \\
$(0 + 1 + 2 + 3 + 4) \hbar \omega $ & ~1.12 & 17.0 & 14.2 & 1.16 \\[3mm]
Unrestricted shell model & & & & \\
$0 \hbar \omega$ & ~0.59 & ~6.9 & ~5.7 & 0.45 \\
$(0 + 1 + 2) \hbar \omega $ & ~0.58 & ~8.4 & ~6.6 & 0.56 \\
$(0 + 1 + 2 + 3 + 4) \hbar \omega $ & (~0.56)\tablenotemark[1] &  
(~7.9)\tablenotemark[1] &  
(~6.3)\tablenotemark[1] &  
(0.53)\tablenotemark[1] \\[3mm]
CRPA\cite{Ko99} & ~0.7(3) & ~8.9(1) & ~6.0(3) & \\[3mm] 
& & $8.9(10)$ \cite{Bo94} & & \\
Expt. & $0.66(14)$ \cite{At97} & $9.1(10)$ \cite{At97}
& $6.1(3)$ \cite{Gi81} & $0.50(3)$ \cite{Aj90} \\
\end{tabular}
\vspace{-5mm}
\tablenotetext[1]{Estimate}
\end{center}
\end{table}

\begin{table}
\begin{center}
\caption{Contribution from each multipole to the inclusive cross section
in a closed-shell RPA + 2p-2h model space. The values on excluding the
$^{12}$N ground-state contribution are given in brackets.
\label{t:tab5}}
\begin{tabular}{clll}
& & & \\[-3mm]
& $(\nu_{\mu},\mu^{-})$ DIF &
$(\nu_e,e^{-})$ DAR &
$\mu$-capture  \\
& $\overline{\sigma} \times 10^{-40}$ cm$^2$ &
$\overline{\sigma} \times 10^{-42}$ cm$^2$ &
$\Lambda_c \times 10^3$ s$^{-1}$ \\
\tableline
Positive parity & & & \\
$0^{+}$ & ~0.11 & ~0.00 & ~0.21 \\
$1^{+}$ & ~2.95~~(~1.79) & 22.52~~(1.66) & 15.43~~(~3.87) \\
$2^{+}$ & ~2.59 & ~0.08 & ~1.36 \\
$3^{+}$ & ~1.39 & ~0.03 & ~0.46 \\
$4^{+}$ & ~0.66 & ~0.00 & ~0.00 \\
$5^{+}$ & ~0.51 & ~0.00 & ~0.00 \\
$6^{+}$ & ~0.04 & ~0.00 & ~0.00 \\
$\geq 7^{+}$ & ~0.04 & ~0.00 & ~0.00 \\
Sum & ~8.29~~(~7.13) & 22.63~~(1.77) & 17.48~~(~5.92) \\[3mm]
Negative parity & & & \\
$0^{-}$ & ~0.07 & ~0.04 & ~2.12 \\
$1^{-}$ & ~3.55 & ~1.90 & 12.25 \\
$2^{-}$ & ~2.91 & ~2.36 & ~7.79 \\
$3^{-}$ & ~1.77 & ~0.00 & ~0.11 \\
$4^{-}$ & ~1.41 & ~0.00 & ~0.07 \\
$5^{-}$ & ~0.17 & ~0.00 & ~0.00 \\
$6^{-}$ & ~0.17 & ~0.00 & ~0.00 \\
$\geq 7^{-}$ & ~0.00 & ~0.00 & ~0.00 \\
Sum & 10.04 & ~4.30 & 22.34 \\[3mm]
Total & 18.33~~(17.17) & 26.93~~(6.07) & 39.82~~(28.26) \\
\end{tabular}
\end{center}
\end{table}

\end{document}